\documentclass []{aa}
\usepackage{psfig} 
\begin{document}
\thesaurus{11.03.4; 11.04.1; 11.12.2; 12.03.3}
\title{A Turn-over in the Galaxy Luminosity Function of the Coma Cluster 
Core?}

\author{C.~Adami \inst{1,2}, 
M.P.~Ulmer \inst{2},
F.~Durret \inst{3,4},
R.C.~Nichol \inst{5},
A.~Mazure \inst{1},
B.P.~Holden \inst{6},
A.K.~Romer \inst{5},
C.~Savine \inst{1}
}

\institute{
IGRAP, Laboratoire d'Astronomie Spatiale, 
Traverse du Siphon, F-13012 Marseille, France  
\and Department of Physics and Astronomy, Northwestern University, Dearborn
Observatory, 2131 Sheridan, 60208-2900 Evanston, USA 
\and Institut d'Astrophysique de Paris, CNRS, 
98 bis Bd Arago, F-75014 Paris, France 
\and DAEC, Observatoire de Paris, Universite Paris VII, CNRS, F-92195 Meudon
Cedex, France
\and Department of Physics, Carnegie Mellon University, 5000 Forbes Avenue,
Pittsburgh, PA 15213, USA
\and Department of Astronomy and Astrophysics, University of Chicago,
5640 S. Ellis Avenue, Chicago, IL 60637, USA
}
\offprints{C.~Adami} 
\date{Received date; accepted date} 

\maketitle 
 
\markboth{A Turn-over in the Galaxy Luminosity Function of the Coma Cluster 
Core?}{} 
 
\begin{abstract} 

Our previous study of the faint end (R$\leq$21.5) of the galaxy
luminosity function (GLF) was based on spectroscopic data in a small
region near the Coma cluster center. In this previous study Adami et al. 
(1998) suggested, with moderate statistical significance, 
that the number of galaxies actually belonging to the cluster was much
smaller than expected. This led us to increase our spectroscopic
sample. Here, we have improved the statistical significance of the
results of the Coma GLF faint end study (R$\leq$22.5) 
by using a sample of 88 redshifts. This includes both
new spectroscopic data and a literature compilation.

The relatively small number of faint galaxies belonging to Coma that
was suggested by Adami et al. (1998) and Secker et al. (1998) has been
confirmed with these new observations. We also confirm that the
color-magnitude relation is not well suited for finding the galaxies
inside the Coma cluster core, close to the center at magnitudes
fainter than R$\sim$19. We show that there is an enhancement in the
Coma line of sight of field galaxies compared to classical field
counts. This can be explained by the contribution of groups and of a
distant $z\sim 0.5$ cluster along the line of sight.

The result is that the Coma GLF appears to turn-over or at least to become 
flat for the faint galaxies. We suggest that this is due to environmental 
effects.

\end{abstract} 
 
\begin{keywords} 
 
{ 
Galaxies: clusters: individual: Coma cluster - 
galaxies: distances and redshifts - 
galaxies: luminosity function, mass function - 
Cosmology: observations
} 
 
\end{keywords}

\section{Introduction}

The galaxy luminosity function (GLF hereafter) is a powerful
cosmological tool (see e.g. the review by Binggeli, Sandage $\&$ 
Tammann 1988). It is a good indicator of the formation history of both 
galaxies and clusters of galaxies, and is also directly related to the mass
distribution function of structures in the Universe (Press $\&$
Schechter 1974). Excluding environmental effects as a first
approximation, hierarchical models predict a mass distribution
characterized by an exponential cut-off beyond a characteristic mass
M$^{*}$ and a power-law for low mass structures. Although a detailed analysis 
presented by Binggeli et al. (1988) shows that individual galaxy types 
actually have a Gaussian distribution, except for the dwarf ellipticals 
(probably the dominant type among the faint Coma galaxies), the GLF 
is now commonly described with a Schechter function
(see e.g. Lumsden et al.  1997 or Rauzy et al. 1998 for recent
statistical applications). It is possible to find a simple linear
relation between the shape of the power spectrum (Fourier transform of
the mass function) and the shape of the GLF (see Lobo et al. 1997).

A determination of the GLF in rich clusters such as the Coma cluster
(see Biviano 1998 for a review including the article by Godwin,
Metcalfe $\&$ Peach, GMP hereafter) is interesting as it can provide
information about the evolution and formation of small galaxies in
rich clusters as well as to place limits to the amount of mass in the
cluster in the form of faint galaxies that might have been missed in
previous surveys. It is also interesting to ask if the luminosity
function varies within projected distance from the core of rich
clusters (cf. Secker et al 1998 and Adami et al 1998: A98 hereafter)
and to determine if the GLF differs in the field versus that in rich
clusters (cf Valotto et al 1997 or Gaidos 1997). Both of these
questions relate to how the environment of rich clusters affects and
is related to galaxy formation and evolution. In order to address some
of these issues several groups have acquired and analyzed deep images
of the Coma cluster to extend the GLF to as faint values as possible
(cf. Bernstein et al, 1995, B95 hereafter, Ulmer et al 1996 and
references therein).

B95 concluded that the luminosity function turned up at the
faint end as also suggested for the case of Virgo by Bingelli et al. (1988). 
Phillipps et al. (1998) confirm this trend with a very deep survey of the
Virgo cluster (${\rm M_R}$ as faint as $-$13.5).
A98 made direct redshift measurements of Coma that tentatively suggested a 
similar turn up, but A98 did not have enough redshifts to draw a 
firm conclusion. In order to redress the issue of too few redshifts, we 
requested and obtained more observing time at the CFHT. With these data and 
a compilation of redshifts from the literature, we now have 88 redshifts that
cover the region surveyed by B95. We can, therefore, make a more definitive 
statement about the 
shape of the luminosity function of Coma at the faint end. We also used our 
data to compare with indirect methods of determining cluster membership, i.e., 
statistical inferences of the number of galaxies in the cluster (B95) and the 
use of the color magnitude relation (e.g. Biviano et al. 1995).

In section 2 of the present paper, we describe our sample of 88 redshifts. In 
section 3, we discuss the implications of these results. Finally, in section 4 
we discuss the GLF.

We will use H$_0$=100 km.s$^{-1}$.Mpc$^{-1}$ and q$_0$=0 throughout
the paper to be coherent with the previous analysis of A98.

\section{The data}

\subsection{Photometry}

We used the photometric catalogue of B95.  All the details
of this catalogue are described in B95 and A98. Briefly, it covered
a 7.5$\times$7.5 arcmin$^2$ area centered on coordinates (12h 57m 17s,
28$^\circ$09'35") (equinox 1950 hereafter). This field is close to the
cluster center, located very near the two giant dominant galaxies of
the Coma cluster. Each galaxy is characterized by its position and its
R and b$_j$ magnitudes.

\subsection{Spectroscopy}

This new spectroscopic follow-up was done during 1 night (11/12 April
1999) using the CFHT/MOS multi-slit spectrograph with the V150 CFHT
grism with a resolution of 10\AA\ px$^{-1}$.  We selected as
potential targets all the galaxies with magnitudes R$\leq$22.5 and
resolution parameter (see definition in B95) between 0.85 and 2.2.
This range of resolution parameters corresponds to the objects 
securely classified as
galaxies. The galaxy selection was confirmed by the very low contamination
rate we found with our spectroscopic observations: less than 5$\%$ of the
targets were found to be stars. The limiting magnitude of
R$\simeq$22.5 is the faintest possible magnitude for a redshift
determination in a non-prohibitive exposure time of about 2 hours.

In A98, we used no blocking filters (wavelength range of
$\simeq$6000\AA\ and about 30 slits per mask). In the current run, we have 
chosen to use a 4400\AA\ wide 
filter (CFHT filter \#4610) which allows a greater number of slits. The 
redshift range is lower but is still acceptable (z in the [0;0.9] range
for both the H$\&$K lines and the G band within our spectral
range). We have shown in A98 that the galaxies in this area of the sky and 
with magnitudes brighter than 22.5 are unlikely to be more distant than 
z=0.9.  Due to bad weather, only 50$\%$ of the night was available for 
spectroscopy. This resulted in a 2 hour exposure with 45 slits (mask 1) 
and a second 40 minute exposure (mask 2) with 40 slits. Therefore, mask 2 
produced a low success rate of redshift determination of
about 15$\%$ (the mask 1 success rate is about 60$\%$).

The spectra were reduced in exactly the same manner as in A98
i.e. we used both ESO MIDAS and Geotran/RVSAO IRAF. We assumed the 
derived redshifts were valid only when the absorption lines
provided a RVSAO correlation coefficient greater than 3 or when there were
more than two well defined and self-consistent emission lines. When both 
emission (at least 2) and absorption lines were detected, the redshift was 
computed with the emission lines. Table 1 summarizes all the information we 
have for the galaxies we used here to derive a GLF (except those of A98).

\subsection{Literature compilation}

In order to increase the number of redshifts in the Coma core area, we
have also searched for velocities available in the literature in the
area surveyed by B95 and with a measured magnitude. The compilation of
A. Biviano (private communication) gives 23 objects. These 23 objects are also 
summarized in Table~1 (objects with a GMP number and without a "*"). 
The colors of these galaxies are from GMP and we have diminished the
b$_{26.75}$-r$_{24.5}$ of GMP by 0.55 in order to match with our
system of magnitudes, but we note this is only an
approximation (the error is about 0.2 magnitude on this number). The
objects from Biviano are significantly brighter than the galaxies we
have observed, but provide complementary data for the bright portion
of the luminosity function.

We note that the objects in Table 1 with a GMP number and a "*" are found both 
in B95 and GMP. 

We had previously described in A98 the data of Secker et
al. (1998). These data are not yet public, but they cover a field very
close to the B95 area. The galaxies of this sample are in the
magnitude range R=[15.5;20.5] and only 4 are in the Coma
cluster. These data are not included in the sample used in this
paper. We simply use the Secker et al. (1998) results for comparison
in section~3.

\begin{table} 
\caption[]{col.1: redshift (with GMP number when the galaxy is in this 
catalog); col.2: redshift error; col.3: R magnitude ; col.4: b$_j$-R; 
col.5: x coordinates (arcmin); col.6: y coordinates (arcmin 
from the center: 12h 57m 17s, 28$^\circ$09'35")}
\begin{flushleft} 
\begin{tabular}{cccccc} 
\hline 
\noalign{\smallskip} 
z & err$_z$ & R & b$_j$-R & x (') & y (') \\ 
\hline 
\noalign{\smallskip} 
0.0162 (3400) & 1 10$^{-4}$ & 13.82 &  1.33  &   3.08  &  -0.10  \\
0.0221 (3201) & 1 10$^{-4}$ & 14.01 &  1.36  &   0.06  &   1.57 \\
0.0207 (3313) & 2 10$^{-4}$ & 14.20 &  1.28  &    1.74  &   -3.61 \\
0.0227 (3423) & 1 10$^{-4}$ & 14.30 &  1.40  &   3.41   &   -2.16 \\
0.0271 (3296) & 1 10$^{-4}$ & 14.38 &  1.41  &   1.53   &   1.25 \\
0.0266 (3178) & 3 10$^{-4}$ & 14.68 &  1.29  &   -0.29  &   -1.76 \\
0.0263 (3068) & 1 10$^{-4}$ & 14.97 &  1.38  &   -2.62 &   2.62 \\
0.0229 (3222) & 3 10$^{-4}$ & 14.97 &  1.20  &   0.53  &   2.30 \\
0.0123 (3262) & 3 10$^{-4}$ & 15.27 &  1.27  &   1.03  &   -1.88 \\
0.0328 (3133) & 3 10$^{-4}$ & 15.73 &  1.37  &   -1.17 &   2.29 \\
0.0214 (3339) & 3 10$^{-4}$ & 16.04 &  1.23  &   2.11  &   -1.35 \\
0.0263 (3126) & 3 10$^{-4}$ & 16.05 &  1.27  &   -1.37 &   -3.20 \\
0.0207 (3205) & 4 10$^{-4}$ & 16.11 &  1.28  &   0.13  &   -1.11 \\
0.0215 (3034) & 2 10$^{-4}$ & 16.56 &  1.15  &   3.26  &   3.25 \\
0.0228 (3376) & 3 10$^{-4}$ & 16.74 &  1.00  &   2.78  &   2.10 \\
0.0564 (3225)* & 1 10$^{-4}$ & 16.75 &  1.23  &   0.54  &   -1.63 \\
0.0153 (3383) & 2 10$^{-4}$ & 17.00 &  1.31  &   2.86  &   -1.50 \\
0.0140 (3340) & 5 10$^{-4}$ & 17.04 &  1.38  &   2.13  &   2.94 \\
0.1782        & 3 10$^{-4}$ & 17.11 &  1.46  &   1.25  &  -0.82 \\
0.0128 (3275) & 1 10$^{-4}$ & 17.16 &  1.02  &   1.16  &   -1.91 \\
0.0204 (3311)* & 4 10$^{-4}$ & 17.54 &  1.23  &   1.69  &   -0.93 \\
0.1715 (3149)* & 5 10$^{-4}$ & 17.41 &  1.67  &  -0.87  &  -2.30 \\ 
0.0187 (3424) & 2 10$^{-4}$ & 17.48 &  1.81  &   3.41  &   3.37 \\
0.1954 (3095)* & 5 10$^{-4}$ & 17.62 &  1.69  &  -2.17  &  -2.17 \\ 
0.0242 (3080)* & 6 10$^{-4}$ & 17.96 &  1.19  &   -2.41  &   1.89 \\
0.0298 (3325) & 5 10$^{-4}$ & 18.16 &  1.45  &   1.94  &   1.20 \\
0.2748 (3150)* & 2 10$^{-4}$ & 18.31 &  1.66  &   -0.87  &  1.02 \\
0.1774 (3284) & 3 10$^{-4}$ & 18.50 &  /     &   1.29  &   -0.88 \\
0.1922        & 3 10$^{-4}$ & 18.55 &  1.27  &  -3.32  &  -1.64 \\
0.6330 (3353)* & 6 10$^{-4}$ & 19.43 &  1.63  &   2.29  &  1.36 \\
0.0273 (3370) & 3 10$^{-4}$ & 19.63 &  1.53  &  2.69  &  -2.07 \\
0.1776 (3137) & 3 10$^{-4}$ & 19.63 &  1.02  &  -1.17  &   2.64 \\ 
0.4680        & 3 10$^{-4}$ & 19.97 &  2.17  &  2.57  &  -0.04 \\
0.3162        & 2 10$^{-4}$ & 20.02 &  1.54  &  -2.07  &  -1.14 \\
0.5060        & 4 10$^{-4}$ & 20.43 &  1.06  &   2.14  &  2.06 \\
0.2438        & 4 10$^{-4}$ & 20.43 &  2.46  &  0.43  &  -1.78 \\ 
0.6142        & 2 10$^{-3}$ & 20.54 &  2.30  &  -1.82  &  -3.53 \\
0.2415        & 3 10$^{-4}$ & 20.55 &  1.23  &   3.67  &  1.41 \\
0.4606        & 3 10$^{-4}$ & 20.66 &  1.33  &   -0.11  &   3.33 \\ 
0.2523        & 4 10$^{-4}$ & 21.08 &  1.12  &  2.97  &  -2.99 \\
0.4229        & 2 10$^{-4}$ & 21.13 &  1.42  &   3.41  &  1.82 \\
0.6262        & 9 10$^{-4}$ & 21.16 &  1.03  &   1.19  &  2.87 \\
0.2368        & 4 10$^{-4}$ & 21.20 &  1.10  &  -3.62  &   0.65 \\
0.4035        & 3 10$^{-4}$ & 21.28 &  1.97  &   0.60  &  2.13 \\
0.3127        & 3 10$^{-4}$ & 21.37 &  0.97  &  -3.34  &  -0.76 \\  
0.6018        & 4 10$^{-4}$ & 21.41 &  2.03  &   3.04  &  2.95 \\
0.4236        & 3 10$^{-4}$ & 21.63 &  1.48  &   1.48  &  2.48 \\ 
0.0934        & 3 10$^{-4}$ & 21.95 &  1.23  &   -0.22  &   0.46 \\
0.1622        & 6 10$^{-4}$ & 22.02 &  1.12  &   1.71  &  -0.56 \\
0.5509        & 2 10$^{-4}$ & 22.13 &  0.89  &   -3.15  &  0.54 \\ 
0.4665        & 3 10$^{-4}$ & 22.23 &  0.58  &   -0.48  &  3.50 \\ 
0.5188        & 4 10$^{-4}$ &   /   &  /     &   5.12  &  0.62 \\
0.2827        & 8 10$^{-4}$ &   /   &  /     &   4.56  &  0.53 \\
0.0926        & 7 10$^{-4}$ &   /   &  /     &  4.38  &  -2.49 \\
0.0619        & 4 10$^{-4}$ &   /   &  /     &  3.89  &  -2.41 \\
0.2134        & 3 10$^{-4}$ &   /   &  /     &   4.44  &  0.55 \\
\noalign{\smallskip} 
\hline	    
\normalsize 
\end{tabular} 
\end{flushleft} 
\label{} 
\end{table}

\subsection{Completeness}

We show in Fig. 1 the distribution of our 88 redshifts along the B95 line of 
sight (A98: 32 z, this paper: 33 z, literature: 23 z). We clearly see the Coma 
cluster at z$\simeq$0.02 and some other structures along the line of sight 
($los$ hereafter) (including the distant cluster detected in A98). The Coma 
cluster comes mainly from the literature.

\begin{figure} 
\vbox 
{\psfig{file=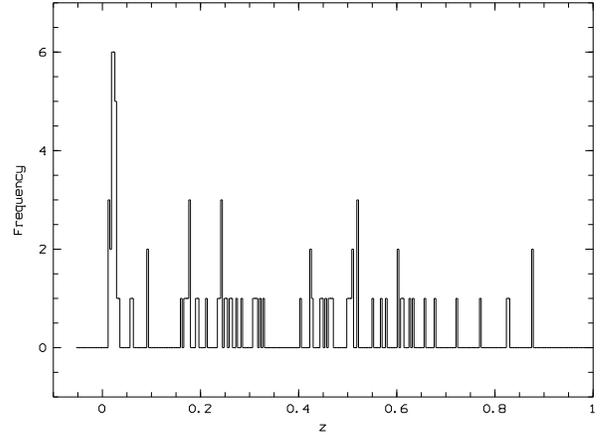,width=9.cm,angle=270}} 
\caption[]{Distribution along the Coma cluster core line of sight of the 88
redshifts of our sample. The bin width is 1000 km.s$^{-1}$.} 
\label{} 
\end{figure}

In order to estimate how complete our sample is in apparent magnitude,
we compare the number of redshifts we produced to the number of available
targets in the magnitude range (R=[16.5;22.23]) in B95 (we ignore the 
very bright galaxies taken from the literature and
not sampled by B95). We find that we covered about 23$\%$ of all the
available galaxies in this magnitude bin. Fig. 2 shows the magnitude
histogram of the galaxies of our sample superimposed on that of all
the B95 galaxies in this magnitude range. We see that our completeness
level drops abruptly above R=21.5. Below this value, our completeness
level is 33$\%$.

We have also tested the spatial representativeness of our sample. We have 
used the same bidimensional Kolmogorov-Smirnov test as in A98 to compare the 
spatial distribution of the galaxies for which we have a redshift and the 
spatial distribution of all the galaxies available in B95 in the range
R=[16.5;22.23]. We find that the probability that the two spatial distributions
statistically differ is less than 0.1$\%$. We conclude, therefore, that our
sample is a statistically representative sample of the Coma luminosity function
and that it is valid to apply corrections to the B95 luminosity function based 
on the data we use here.

Finally, we note that we are not likely to be strongly biased by
non-detected very low surface brightness galaxies. Even the lowest
surface brightness galaxies of our sample (described in Fig. 5 of
Ulmer et al. 1996) with magnitudes brighter than R=22 (${\rm M_B}\simeq -12$ 
in
Fig. 5 of Ulmer et al. 1996) have a central surface brightness brighter
than 26.5 mag.arcsec$^{-2}$. On the same figure, the galaxies studied
by Ferguson $\&$ Binggeli (1994) for a similar magnitude range have
the same central surface brightness. We are, therefore, not biased more than 
Ferguson $\&$ Binggeli (1994) are.

\begin{figure} 
\vbox 
{\psfig{file=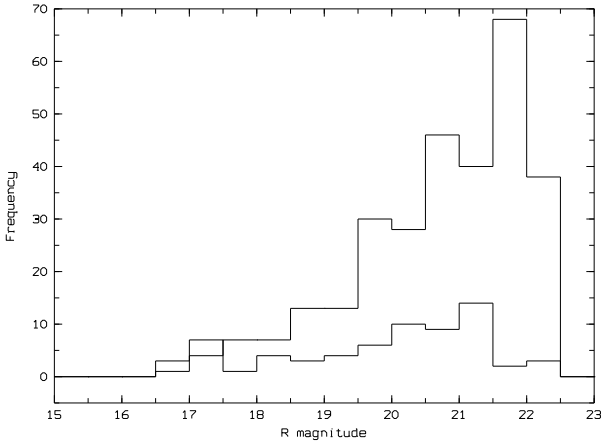,width=9.cm,angle=270}} 
\caption[]{Magnitude histograms for all the B95 and literature galaxies 
for which we have a measured redshift.} 
\label{} 
\end{figure}

\section{Implications of the results}

\subsection{The Color-Magnitude Relation (CMR)}

We concluded in A98 that the use of the CMR to find the galaxies
inside the Coma cluster for the faint magnitudes (typically fainter
than R=19) was inefficient. We confirm this statement here. By using
all the galaxies for which we have a b$_j -$R color, we see that some
of the galaxies beyond the cluster are included in the envelope
expected for Coma (see Fig. 3), while for the
bright galaxies the relation is very good. This contamination is
likely due to distant intrinsically blue galaxies with a K-dimming
that makes them mimic Coma cluster members.  Yet, the two galaxies
detected inside the Coma cluster in our spectrocopic survey (the two
faintest filled circles, see Table 1 for relative coordinates) are
redder than this envelope (they are not low surface brightness
galaxies and they have a mean surface brightness brighter than 23.5
mag.arcsec$^{-2}$). We note however that the redshift of the first of
these two galaxies is based on only one emission line (see A98). In
order to check this tendency with a larger sample, we have plotted on
Fig. 3 the running mean (thin solid line with a running window of 10
galaxies) of the mean color of all the galaxies of the GMP catalog
belonging to the Coma cluster (with a known redshift), not only in the
small field of B95 (redshifts from Biviano et al. 1995 and included in
[3300;10500]km.s$^{-1}$). Except for a few outliers (a few percent),
the Coma galaxies brighter than R=17.25 are in good agreement with the
mean position of the CMR. For the fainter GMP galaxies the agreement
is also good globally, but the percentage of red outliers increases
(25$\%$) in the data (according to the envelope of the CMR, see Mazure
et al. 1988).  Finally, the two objects we observed are clearly redder
than the CMR envelope.

This is consistent with the suggestion of A98 that there exist high
metallicity dwarf galaxies in the Coma cluster core. We do not
generalize this hypothesis to all the dwarf galaxies because our
statistics are very poor.  Different hypotheses can explain, however,
the existence of red dwarfs. For example, we can assume that the
intra-cluster medium confines these faint galaxies (when they are not
destroyed by tidal forces), thereby keeping a high fraction of heavy
elements.

\begin{figure} 
\vbox 
{\psfig{file=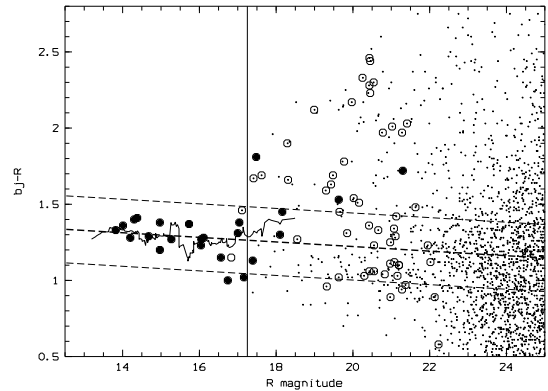,width=8.cm,angle=270}} 
\caption[]{Colour--Magnitude Relation (CMR) for the galaxies between magnitudes
R=13 and 25. The galaxies observed spectroscopically are circled and the 
galaxies belonging to the Coma cluster are the filled circles (the faintest
ones are not from the literature compilation). The mean position of the CMR 
envelope of the cluster is overplotted as a thick dotted line. We note that
the two circles without a point inside are two galaxies classified as stars
by B95. We have overplotted the running mean of the GMP galaxies belonging to 
the Coma cluster (thin solid line inside the CMR envelope). We diminished
the b$_{26.75}$-r$_{24.5}$ of GMP by 0.55 and b$_{26.75}$ by 1.5 in order to 
match with our system of magnitudes. The thin vertical line shows the
magnitude R=17.25.
} 
\label{} 
\end{figure}

\subsection{Projection effects}

In order to reconstruct the GLF of the Coma cluster core, we need to
know the number of galaxies per magnitude bin belonging to the cluster. One 
of the classical methods to obtain this information is to observe 
comparison fields (as in B95) and to make differential counts between these 
fields and the cluster field. We plot in Fig. 4 the percentage of galaxies 
along the $los$ which are in the Coma cluster according to the B95 field counts
(dotted line with the $\pm$ 1$\sigma$ error bars). The error bars are the rms
deviation on the B95 field counts and are described in B95 section 3.1.
We also plot on the same figure this percentage according to the redshifts we 
have (solid line). This curve was computed using a running mean. This running 
mean was created with an average redshift in each bin of 10 galaxies. The bins 
were generated by shifting from low to high redshift by just one galaxy.
We used this running mean in order to reduce the noise due to our 
small sample. We see that the spectroscopic and the B95 field counts are 
inconsistent at the 1$\sigma$ level for R fainter than 18.5. We note that 
for the two faintest bins, the B95 error bars are large.

We also plot the same percentage, but with all the GMP galaxies with a 
redshift and not only in the B95 field. For magnitudes brighter than R=17.75,
the GMP spectroscopic counts are higher than ours indicating a higher
percentage of galaxies belonging to the Coma cluster if we consider the
whole cluster instead of the small B95 field. They agree well with the 
field counts of B95. For the faintest magnitudes of GMP, the GMP counts agree
better with our counts than with the B95 counts.

We also note that our estimates are the same as those of
Secker et al. (1998). These authors have measured 17 redshifts in the
magnitude range [15.5;20.5] and find only 4 galaxies in the Coma
cluster; this gives a percentage of 24$\%$. If we make the same
estimate for the same magnitude bin with the data of this paper, we find 
that 25$\%$ of the galaxies are in the Coma cluster.

\begin{figure*} 
\vbox 
{\psfig{file=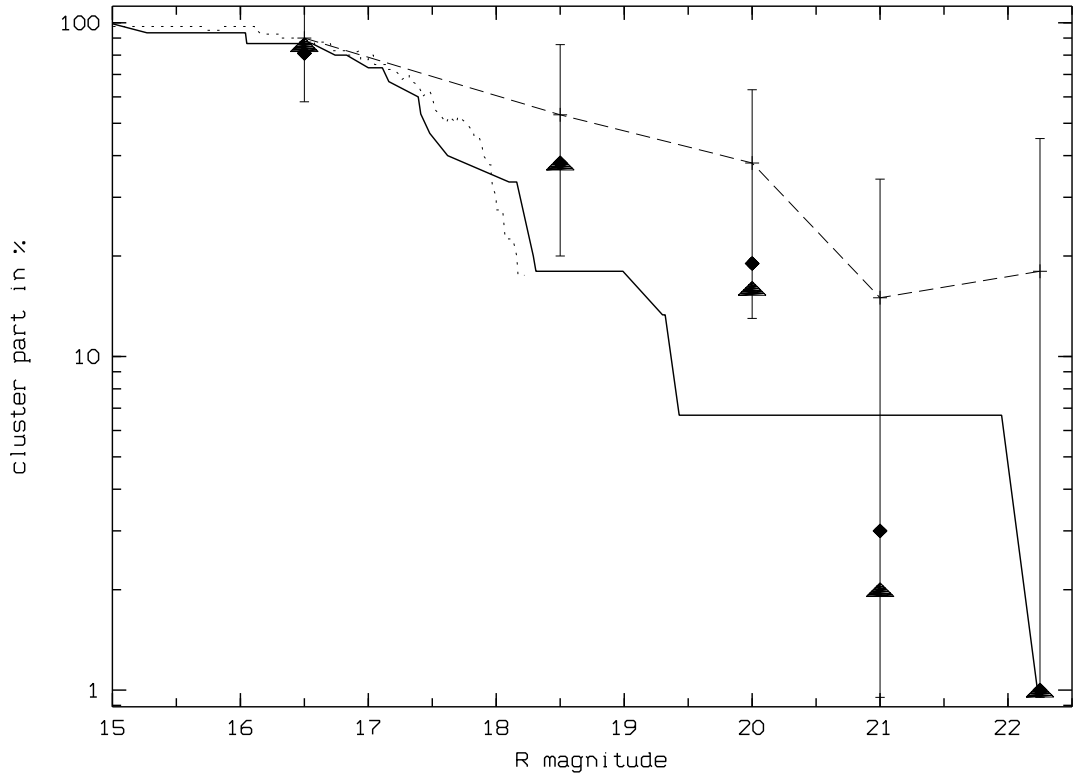,width=16.cm,angle=270}} 
\caption[]{Percentages of galaxies seen along the $los$ which are
in the Coma cluster. Dashed line with 1$\sigma$ error bars: based on the B95 
field counts; solid line: from spectroscopy, this curve was computed 
with a running mean with 10 galaxies in the running window; short dashed
line: from the GMP catalogue, this curve was computed 
with a running mean with 40 galaxies in the running window;
filled diamonds and triangles: based on the B95 field counts but corrected 
for the effect of the distant cluster at z$\simeq$ 0.5 and for the effect of 
groups along the $los$ (see text for a description of these corrections). The 
triangles are the counts corrected with a group luminosity function and the 
diamonds are the counts corrected with a cluster luminosity function for the 
groups along the $los$. We note that these diamonds and triangles are 
overlapping in the first two bins and in the last bin.} 
\label{} 
\end{figure*}

\subsection{Why do we have so few galaxies belonging to the Coma cluster?}

The expected absolute magnitudes for typical cluster galaxies range from 
${\rm M_R}\simeq -23$ to ${\rm M_R} \simeq -11$
(see e.g. Patterson $\&$ Thuan 1996). At the Coma cluster
redshift, we should expect a significant number of galaxies for
apparent magnitudes R close to 23--24. The observational evidence is that
we do not detect such a significant number in the Coma cluster
core. Secker et al. (1998) and A98 have suggested that the faint galaxies 
are disrupted by
the giant ellipticals in the center of the clusters. This explanation
is theoretically supported by the simulations of Merritt (1984). He
predicted a tidal radius (i.e. the lowest possible value for the size
of a non tidally disrupted galaxy) of about 15 kpc in the core of
clusters. The size of dwarf galaxies is significantly lower. More
recent simulations by Moore et al. (1998) also show a similar
scenario for the disruption of faint galaxies.

If these galaxies are really destroyed by tidal forces, their
luminosities can be transferred from the ``galaxy" matter phase to the
``diffuse component" phase. Gregg \& West (1998) propose such an
explanation for the Coma cluster, estimating that 20\% of the diffuse
luminosity of a cD galaxy may come from these disruptions. More
quantitatively, B95 show that the diffuse light profile is strongly
peaked at the center of the Coma cluster and Ulmer et al. (1996)
reported that the color of the diffuse light is b$_j -$R=1.2$\pm$0.3 ,
similar to the expected color for dwarf galaxies inside the Coma
cluster (slightly redder). These two facts support the
disruption hypothesis. Some faint low surface brightness galaxies can
also be produced by this process (Ulmer et al. 1996) to fit with our
explanation of the death of faint galaxies in the Coma core; however,
the overall distruction rate must be faster than the production rate.

There are several other explanations besides tidal disruption for
the scarity of dwarfs in the core of rich clusters. For example, as suggested
by the referee, the formation of faint dwarf galaxies could have been 
suppressed due to heating of the baryons in the relatively high density
fluctuation that caused Coma to form. In this case, such high temperatures
could have prevented faint dwarfs from forming.

\subsection{Why do we see so many galaxies along the Coma cluster core line
of sight?}

Five comparison fields (B95) were observed under identical conditions
(same telescope, same instrument, same filter) as for the B95 Coma
field.  They were selected at random and chosen to be free of bright
stars or galaxies. The flat-fielding and image combining techniques
were the same as for the B95 Coma field. All the details are given in
B95. The errors on the B95 counts come from a calculation based on the
variations of the counts in these five fields.

The question is why do we see a high number of faint galaxies (in apparent magnitude) along the Coma cluster core $los$ compared to these standard empty 
fields? The only explanation we have is an accidental overdensity of structures 
along this $los$ which artificially increases the counts.

One of these structures has been detected by A98: a cluster at
z$\simeq$0.5.  Moreover, using the compilation by A. Biviano of all
the literature redshifts for the Coma cluster $los$ (not only the B95
$los$), we have detected 8 other structures beyond Coma. The detection
in the redshift space is based on the method used for the ENACS
clusters of galaxies, described in Katgert et al. (1996). Each
structure is sampled with more than 5 redshifts and the estimation of
the velocity dispersion $\sigma$ (Biweight Estimators) ranges from 170
to 500 km s$^{-1}$. These values are consistent with groups of galaxies
(see e.g. Barton et al. 1998).

In order to estimate the contribution of these structures to
the counts, we have first searched for the groups that overlap
with the B95 field $los$. Assuming that these eight groups are all
potentially compact groups embedded in a dense environment, we 
assumed that their contribution is significant up to 1 Mpc in radial 
projection from the center of the mass concentration, as suggested by Barton 
et al. (1998). Fig. 5 shows the
location of the B95 field and the 2 Mpc $\times$ 2 Mpc boxes for all these
groups. We see that we have three overlapping groups, that contribute
7~$\%$, 4~$\%$ and 20~$\%$ of their area to the B95 $los$. They are described
in Table 2.

\begin{table} 
\caption[]{Description of the three groups and of the cluster along the line 
of sight. Line (1): mean redshift ; line (2): estimate of the velocity 
dispersion ; line (3): estimate of the contribution in term of number of 
galaxies along the line of sight down to R=16.5 ; line (4): estimate of the 
contribution in term of number of galaxies along the line of sight in R=[16.5,
19.5] ; line (5): estimate of the contribution in term of number of galaxies
along the line of sight in R=[19.5,20.5] ; line (6): estimate of the 
contribution in term of number of galaxies along the line of sight in 
R=[20.5,21.5] ; line (7): estimate of the contribution in term of number of 
galaxies along the line of sight in R=[21.5,23]. For the 3 groups, the first 
number of galaxies is the contribution assuming the Rauzy et al. luminosity 
function, and the second number is if we assume the Muriel et al. luminosity 
function.}
\begin{flushleft} 
\begin{tabular}{ccccc} 
\hline 
\noalign{\smallskip} 
 & Group1 & Group2 & Group3 & dist. cluster\\ 
\hline 
\noalign{\smallskip} 
Mean z & 0.0907 & 0.0970 & 0.2753 & $\sim$0.5 \\
$\sigma$ (km.s$^{-1}$) & 510 & 235 & 170 & 660 \\
R$\leq$16.5 & 0 or 0 & 0 or 0 & 1 or 1 & 0 \\
16.5$\leq$R$\leq$19.5 & 1 or 1 & 0 or 0 & 3 or 4 & 1 \\
19.5$\leq$R$\leq$20.5 & 1 or 1 & 0 or 0 & 6 or 6 & 6 \\
20.5$\leq$R$\leq$21.5 & 2 or 1 & 0 or 0 & 11 or 7 & 17 \\
21.5$\leq$R$\leq$23.0 & 4 or 2 & 1 or 0 & 23 or 10 & 60 \\
\noalign{\smallskip} 
\hline	    
\normalsize 
\end{tabular} 
\end{flushleft} 
\label{} 
\end{table}

\begin{figure} 
\vbox 
{\psfig{file=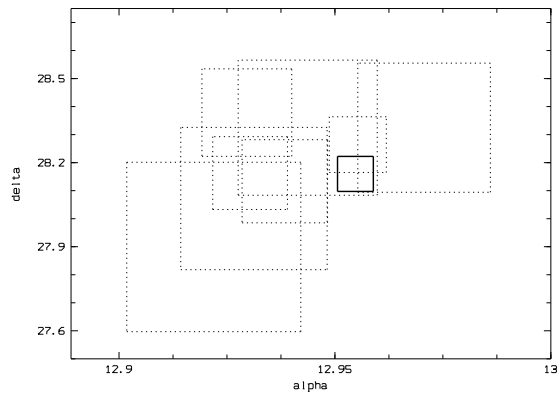,width=8.0cm,angle=270}} 
\caption[]{Location of the B95 field (thick box) and of the 2 Mpc 
side boxes (dashed lines) for all the groups along the Coma line of sight.} 
\label{} 
\end{figure}

We now estimate the contribution of these three overlapping groups and
of the distant cluster (detected along the $los$ in A98) to the counts. 
In order to find an upper limit for the
effect, we assume that the distant cluster is a very rich one,
with the richness of A3112 (the richest cluster in the ENACS sample
of Rauzy et al. 1998) and we assume the luminosity function of Rauzy
et al. (1998): $\alpha =-1.5$ and M$_{bj}$$^{*}$$=-19.91$.

For the groups, we have used the relation $\sigma$=5.28$\times$N+106
given by Barton et al. (1998) to evaluate their richness N to a given
limiting magnitude. For their luminosity function, we have both used
the Rauzy et al. (1998) cluster function and the Muriel et al. (1998)
group function ($\alpha =-1.0$ and M$_{bj}$$^{*}=-19.6$).

We have then predicted the number of galaxies along the Coma cluster
core $los$ coming from these groups and from the distant
cluster. Subtracting these numbers to the B95 field counts, we plot
the corrected values in Fig. 4 as filled symbols: triangles when we
used the Rauzy et al. (1998) function for the groups and squares when
we used the Muriel et al. (1998) function. We see that we are able to
produce values in agreement with those derived spectroscopically in the 
two faintest bins. With these corrections, the [17.5;19.5] and [19.5;20.5] 
bins also become consistent with the spectroscopic results. We
therefore conclude that the presence of groups and of the distant
cluster along the line of sight can explain the difference between the B95 and 
the spectroscopic counts. However, we have used an extreme 
hypothesis for the richness of the groups and of the distant cluster.

We note that such superpositions are found in the ENACS clusters as
shown in Katgert et al. (1996). However, this study also shows that
the contamination level spreads over a wide range and that all the
clusters are not subject to such superposition effects.

\section{The GLF}

We show in Fig. 6 the GLF of B95 (rebinned to conform better to our
sample) and the GLF constructed with our own spectroscopic counts. We
produced the brightest two points of our GLF by using the Lobo et
al. (1997) photometric data included in the B95 field and by applying
the spectroscopic percentages of galaxies belonging to the cluster
(see Fig. 4: $\sim$100$\%$ of the galaxies in the Coma cluster). We
produced the last five points of our GLF by using the spectroscopic
counts computed with the B95 data and by applying the spectroscopic
percentages of galaxies belonging to the cluster (instead of the field
counts). While B95 quotes a best fit to the R$\simeq$15-24 GLF of
-1.42$\pm$0.05, the GLF in the R$\simeq$15-22 range is consistent with
a Schechter function with a slope of $\alpha$=-1. After our
corrections are applied, the GLF seems to be flat after R = 13 and may
actually be begin to turn over at R$\simeq$17 or higher. We note,
however, that this turn-over is not statistically significant given
the size of the error bars and the turn-over (if there is one) could
take place anywhere between R$\simeq 17$ and R$ \simeq 21$. These
errors are difficult to estimate due to our limited number of
redshifts, but we have assumed that they have the same values as those
of B95. Given these uncertainties, we conclude that the GLF is most likely 
flat (flatter than deduced by B95), though a moderate rise or decrease are 
not excluded. The GLF slope (using dwarf
galaxies) is the same as in Loveday et al. (1992:
$\alpha$=0.20$\pm$0.35 in a magnitude range b$_j$=[-22;-15]) for
``normal" elliptical field GLF and we have overplotted this function
in Fig. 6. Their turn-over, however, begins around b$_j =-19.75$
(R$\simeq -21.25$), sooner than for our GLF as we can see on
Fig. 6. We therefore conclude that the GLF of the Coma cluster core is
not in agreement with the field elliptical GLF. However, the Coma
cluster core GLF is also significantly different from the GLF for the
whole Coma cluster of Lobo et al.  (1997) or Smith et al. (1997) (the
faint ends of the GLF are respectively at ${\rm M_R} \simeq -14$ and
${\rm M_R} \simeq -14.5$), or of Phillipps et al. (1998) for the Virgo
cluster. These authors find a slope for the GLF of $\alpha \simeq -1.8
$ or greater. The conclusion is that the Coma cluster core GLF has a
peculiar shape that was probably induced by environmental
effects. This shape is flat for the faint magnitudes, similarly to
what is observed in the field (e.g. Lin et al. 1996 without
morphological distinctions). However, we stress here that any general
conclusion concerning the similarity between the Coma inner core GLF
and the GLF in the field is not valid because we are very likely to
have strong environmental effects in the inner core of the Coma
cluster.

A lack of faint galaxies is likely to be due in a large part
to the tidal forces created by the two giant dominant galaxies, and
an interesting question is: what is the variation of the
efficiency of this disruption versus the distance from the center? 
Do environmental effects dominate the shape of the faint end of the GLF at 
larger radii? If this phenomenon is localized to the few central 100 kpc of 
the cluster, the effect on the global GLF of the entire cluster will be 
limited to a correction of a few percent. Another question to address is 
do we see the same tendency in the clusters where we have only one giant 
dominant galaxy or none? In order to answer to these questions, more 
spectroscopic data are required with at least a similar magnitude depth
over a larger area of the Coma cluster and other clusters need to be
studied as well.

\begin{figure} 
\vbox 
{\psfig{file=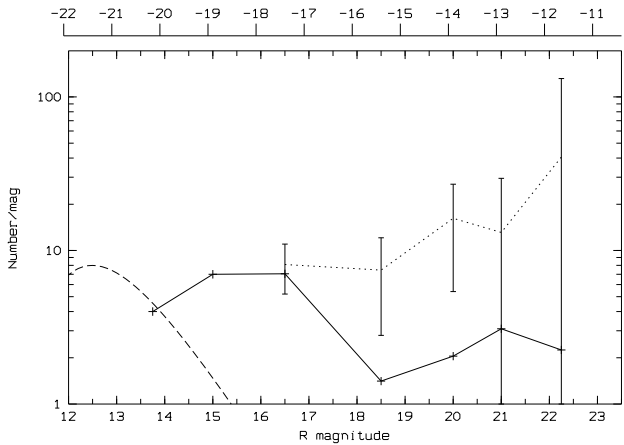,width=9.5cm,angle=270}} 
\caption[]{Galaxy Luminosity Function constructed with the B95 counts 
(dashed line with 1$\sigma$ error bars) and with the spectroscopic counts 
(solid line). The long-dashed line is the GLF of Loveday et al. (1992) for
the field elliptical galaxies. This GLF has been arbitrary normalized at eight 
galaxies for R = 12.5. We have used a distance modulus of 33.90 for the Coma 
cluster and we give the corresponding R band absolute magnitudes at the top of 
the figure. The number of galaxies per unit of magnitude is in the B95 field.} 
\label{} 
\end{figure}

\section{Conclusion}

We have used a sample of 88 redshifts along the B95 $los$ to investigate the 
GLF in the Coma cluster
core. We find that the only two faint Coma galaxies in our sample (R fainter 
than 19), are redder than the envelope of the CMR in the Coma cluster core. The
explanation could be the confinement of the metal-rich gas of these galaxies by 
the intra-cluster medium. Alternatively, these two galaxies can also have 
formed from the recent disruption of old infalling galaxies.

We find that there are only a few galaxies in the Coma cluster with magnitudes
fainter than R=18.5. The field counts of B95 predicted significantly more
galaxies in the Coma cluster velocity field, but this can be explained by 
accidental superposition effects of 3 groups and 1 distant cluster along the
B95 field $los$.

Using the spectroscopic counts to correct the GLF of B95, we find a nearly
flat shape for this function for the magnitudes fainter than R=17 and 
possibly a turn-over in the GLF between R$\simeq$17 and 21. This is likely due 
to environmental effects leading to the disruption of most of the faint 
galaxies by the two dominant galaxies of the Coma cluster core.

\begin{acknowledgements}

{AC thanks the staff of the Dearborn Observatory for their hospitality 
during his postdoctoral fellowship. BH would like to acknowledge support from 
the following: NSF AST-9256606, NASA grant NAG5-3202, NASA GO-06838.01-95A, 
and the Center for Astrophysical Research in Antartica, a National Science
Foundation Science and Technology Center. The authors thank the referee,
N. Trentham, for very useful comments.}

\end{acknowledgements}

\vfill 
\end{document}